\documentclass[11pt,a4paper]{article}

\usepackage[utf8]{inputenc}
\usepackage[T1]{fontenc}
\usepackage{amsmath,amssymb}
\usepackage{graphicx}
\usepackage{booktabs}
\usepackage{hyperref}
\usepackage[margin=2.1cm]{geometry}
\usepackage{xcolor}
\usepackage{algorithmic}
\usepackage{algorithm}
\usepackage{multirow}
\usepackage{enumitem}
\usepackage{caption}
\usepackage{subcaption}
\usepackage{fancyhdr}

\title{No\=esis: Deterministic-First Retrieval with Two-Tier Context Hydration\\for Factuality-Critical Queries on Small Local Models\thanks{Patent pending. Application No.~102026000023146; Application No.~IT202500035167. This paper is the query-plane counterpart of our companion work on the No\=esis architecture~\cite{cogotti2026noesis}.}}

\author{Nicola Cogotti\\Alpha Cogs\\
\texttt{nicola.cogotti@alphacogs.co.uk}}

\date{September 2026}

\begin{document}
\maketitle

\vspace{-1em}
\begin{center}
\small\textit{Patent pending: Application No.~102026000023146; Application No.~IT202500035167.}
\end{center}
\vspace{-0.5em}

\begin{abstract}
\sloppy
Retrieval-Augmented Generation (RAG) is increasingly deployed on small, local language models for latency, cost, and data sovereignty. Yet answer quality remains tightly coupled to model scale: sub-7B models fail to utilize retrieved context effectively, and standard RAG can be a net negative at this scale~\cite{pandey2026small}. We argue this coupling is an artifact of the query interface, not of model capacity. On factuality-critical queries---exact values, cross-source alignment, multi-source attribution---the dominant failure is not retrieval miss but confabulation: models fabricate numbers and timestamps even when correct evidence is in context.

We present \textbf{No\=esis}~\cite{cogotti2026noesis}, the deterministic-first query plane of the No\=esis architecture, which makes every deterministic judgment before generation. Its mechanisms are a direct consequence of the ingestion architecture---the subject of a separate patent application: (a)~a producer-side fact layer rendering precomputed metric facts verbatim without ranking; (b)~positional addressing with deterministic cross-source alignment, resolved ahead of query time at zero LLM cost; (c)~provenance scoping as an attribution constraint with multi-tier named-reference routing; and (d)~two-tier context with model-triggered verbatim hydration.

Across four ablations, a 2B-parameter model achieves parity with a 35B model on factual integrity for factuality-critical questions (exact values in all runs; zero confabulated numbers on absent-entity traps); structured retrieval outperforms flat RAG by $+11.4$ points at 2B; skeleton-only context preserves quantitative answers at $-20/30\%$ prompt size; and hydration recovers verbatim narrative in $\sim$8\,s versus $\sim$29\,s for eager context. Two properties matter for regulated domains: each factuality-critical query resolves in a single generation call, and every reported value is traceable to its exact source and position by construction.
\end{abstract}

\section{Introduction}
\label{sec:intro}

Retrieval-Augmented Generation grounds large language model (LLM) responses in external corpora~\cite{lewis2020rag}. When the generator is a small local model---the regime of on-premises, latency-sensitive, and data-sovereign deployments---two problems dominate. We describe them, then present the query plane that addresses both.

\paragraph{Model-Scale Coupling.} Deploying RAG on small language models (SLMs) keeps inference fast, cheap, and inside the organization's network perimeter~\cite{subramanian2026b1ade}. A recent controlled study across scales from 360M to 8B parameters carefully characterizes a limit of this regime: below 7B the main limitation of RAG is not retrieval quality but \emph{context utilization}---even when the retrieved passage is guaranteed to contain the answer, small models often fail to use it reliably, and adding retrieval context under standard conditions can be a net negative at this scale~\cite{pandey2026small}. A common reading of such results is that better answers require bigger models.

\paragraph{Confabulation on Factuality-Critical Queries.} We observe a failure mode we consider more severe for production use. On factuality-critical queries---questions whose answer is an exact value at an exact position (a peak audience share of 37.5\% at second 146; the minimum of a series; the difference between two programs)---the dominant error is not retrieval miss but \emph{confabulation}: the model emits plausible numbers and timestamps that are absent from the corpus, even when the correct evidence is present in context. This query class is central to any factuality-critical domain---audience metrics, scheduling and rights in media; dosages and lab values in healthcare; figures and citations in finance and legal---and it is the one where a wrong number is not merely unhelpful but actively harmful: a decision made on a fabricated value (an editorial call on a peak share, a clinical reading, a financial figure) is worse than no answer at all. We evaluate on broadcast media as one such domain, chosen for its unambiguous ground truth; the mechanisms assume no particular subject matter.

\paragraph{Context Bloat under Eager Retrieval.} A second problem compounds the first. Standard RAG assembles context eagerly: top-$k$ prose chunks are retrieved and injected regardless of whether the question needs them, a pattern associated with context overload and indiscriminate retrieval in agentic settings~\cite{mishra2026sok}. For quantitative questions, prose is not merely redundant---it competes for attention with the few deterministic values the answer depends on, and it inflates prompt size exactly where small models degrade most. The field's response has been to add machinery: learned retrieval triggers~\cite{jiang2023flare,asai2024selfrag}, corrective evaluators~\cite{yan2024crag}, compression models~\cite{jiang2023llmlingua,xu2024recomp}. We take the opposite direction: we remove from the model every judgment that the infrastructure can make deterministically, and we let the model \emph{request}---rather than receive---the prose it actually needs.

\paragraph{Positioning.} This paper is the query-plane counterpart of our companion work on the No\=esis architecture~\cite{cogotti2026noesis}, which presented the ingestion side of the system (bidirectional graph traversal, adaptive parallelism, model quantization, cross-KB semantic discovery). The mechanisms reported here are a direct consequence of that architecture. We make no claim that any single mechanism is new in isolation; several have close prior art (\S\ref{sec:related}). Our contribution is their conjunction into a deterministic-first query interface---made explicit as an assembly principle we call \emph{epistemic tiering} (\S\ref{sec:tiering})---and its measured effect on small-model answer quality.

\medskip
\noindent\textbf{Contributions.} We present No\=esis, the fully implemented and validated query plane of the No\=esis system:
\begin{enumerate}[leftmargin=*]
    \item A \textbf{producer-side deterministic fact layer}: metric facts computed at ingestion time, materialized as graph nodes, rendered verbatim in context without any threshold level (\S\ref{sec:facts});
    \item \textbf{Positional addressing with deterministic cross-source alignment}: content made addressable by position along an ordered attribute, related to positional observations ahead of query time at zero LLM cost (\S\ref{sec:axis});
    \item \textbf{Provenance scoping as an attribution constraint} and multi-tier named-reference routing: no lossy signal can veto a named reference; scoping bounds attribution, never data access (\S\ref{sec:scoping});
    \item \textbf{Two-tier context with model-triggered verbatim hydration}: a structural skeleton without prose, completed on demand by inventory-addressed fetches the model itself requests---training-free, no control tokens, no constrained decoding (\S\ref{sec:hydration});
    \item \textbf{Epistemic tiering and address-bounded assembly}: context is assembled by the mechanism that selected each piece of evidence---computed facts, addressed spans, discovered chunks---each with its own section, provenance, and scope; once deterministic addressing resolves a file set, semantic fill operates inside it (a bilevel structure), with an escape valve so the constraint bounds noise but never coverage (\S\ref{sec:tiering});
    \item A \textbf{four-arm ablation study (A--D)} demonstrating that, for factuality-critical queries over producer-structured corpora, a 2B local model matches a 35B model when the deterministic infrastructure is present, and degrades gracefully when it is removed (\S\ref{sec:eval}).
\end{enumerate}

\section{Architecture Overview}
\label{sec:arch}

No\=esis inherits from the companion architecture~\cite{cogotti2026noesis} a decoupled producer/query design. The \emph{producer side} (ingestion) transforms heterogeneous sources into a structured knowledge base with per-source provenance; its internals are the subject of the companion work and are out of scope here. The \emph{query side} (this paper) activates that structure for a question and assembles the context the model will see.

The organizing principle is \textbf{producer/query decoupling}: each knowledge base declares what it owns, and the query plane consumes those declarations without interpreting formats. The graph acts as a first-class navigator that answers \emph{where} information lives; embeddings fetch at that location. This inverts the common RAG priority order, in which vector similarity decides and structure (if present) decorates.

Table~\ref{tab:mechanisms} summarizes the four mechanisms of this paper, where each acts, and what it guarantees.

\begin{table}[t]
\centering
\small
\caption{Mechanisms of No\=esis. ``Guarantee'' states the property that holds by construction, not by score.}
\label{tab:mechanisms}
\begin{tabular}{@{}p{4.2cm}p{3.6cm}p{7.0cm}@{}}
\toprule
\textbf{Mechanism} & \textbf{Acts at} & \textbf{Guarantee by construction} \\
\midrule
Deterministic fact layer (\S\ref{sec:facts}) & Producer (ingestion) & Exact metric values are always in context; never demoted or filtered by any score \\
Positional addressing + cross-source alignment (\S\ref{sec:axis}) & Producer (index time) & Positional queries answerable without LLM; cross-source ``what was happening when'' is deterministic \\
Provenance scoping + named-reference routing (\S\ref{sec:scoping}) & Query (activation) & A named reference deterministically reaches its owning source/KB; attribution never crosses sources \\
Two-tier hydration (\S\ref{sec:hydration}) & Query (context assembly) & Prose cost is paid only when the model requests it; fetches are bounded to the named file and validated against the KB inventory \\
Epistemic tiering + address-bounded fill (\S\ref{sec:tiering}) & Query (context assembly) & Addressed content never competes with ranked evidence for attention; once an address resolves, semantic fill stays inside the addressed files; no resolvable address $\Rightarrow$ byte-identical unconstrained assembly \\
\bottomrule
\end{tabular}
\end{table}

\section{The Deterministic Fact Layer}
\label{sec:facts}

\paragraph{Problem Statement.} Large language models are unreliable at arithmetic over raw data series, and small models fail systematically. A question such as ``what was the peak share of program $X$, and when did it occur?'' requires scanning hundreds of rows to locate a maximum and its position; in standard RAG the model receives those rows as prose chunks and must perform the computation itself. Every semantic judgment left to the model is a failure mode that has not been eliminated---and for factuality-critical queries, an arithmetic error is indistinguishable from a confident answer.

\paragraph{Design.} No\=esis moves this computation to the producer side. At ingestion time, for each metric series declared by the knowledge base, extrema are computed together with their positions and materialized as deterministic facts in the knowledge base, related to the content at the same position (\S\ref{sec:axis}). At query time, when activation fires on an entity or coordinate, its deterministic facts are rendered verbatim into a dedicated section placed at the top of the context---before any prose. Two properties matter. First, \textbf{deterministic data has no threshold level}: facts are not scored, ranked, or filtered; they either exist in the KB or they do not, and when present they always reach the model. Second, \textbf{vocabulary is owned by the producer}: surface variants of a metric name are resolved by producer-declared vocabulary, so they map to the same series without model interpretation.

\paragraph{Measured Results.} In the ablation study (\S\ref{sec:eval}), all quantitative questions (peak value with position; cross-program comparison with explicit difference; minimum value with position) were answered with exact values in every run, by both models and under both context arms---28/28 quantitative runs correct, including a 2B model that explicitly showed its arithmetic (``higher by 1.0 percentage point ($37.5 - 36.5$)'').

\section{Positional Addressing and Cross-Source Alignment}
\label{sec:axis}

\paragraph{Problem Statement.} Much content is positional: a broadcast has seconds, an episode has scenes, a document has ordered sections. Questions such as ``what was happening at position $p$?'' or ``which source was active when $X$ peaked?'' require interval-level grounding across \emph{different sources} (for example a metric table and a transcript). Embeddings cannot answer interval queries; LLM-based alignment is confabulation-prone, non-reproducible, and pays model cost at query time for what is in fact a deterministic join.

\paragraph{Design.} Content is made positionally addressable along an ordered attribute declared by the knowledge base. The mechanism is \textbf{generic}: it carries no domain-specific semantics, so the same addressing works for any ordered attribute; format knowledge is owned by the producer, not by the query plane. Cross-source alignment---relating positional observations to the content occupying the same interval---is resolved deterministically ahead of query time, so ``what was happening when $X$ peaked'' is answered by traversal over precomputed relations, not by model reasoning. The query-time fidelity rule rendered into context is likewise generic: it instructs the model that positional claims must come from the addressed attributes, without naming any particular one---universality across formats follows from keeping the query plane free of domain vocabulary.

\paragraph{Measured Results.} Open-ended questions spanning both programs in a KB were answered with correct per-program metrics and positions in all runs (\S\ref{sec:eval}, S6), and no positional confabulation occurred anywhere in the battery: when a position was asked, it came from the addressed attributes; when it was not available, the model said so.

\section{Provenance Scoping and Named-Reference Routing}
\label{sec:scoping}

\paragraph{Problem Statement.} In multi-entity corpora, entities recur across files and episodes---similar program names, characters with identical labels in different shows. Two failure modes follow. \emph{Cross-file confusion}: the model attributes data from source $B$ to the named source $A$. And \emph{lossy-signal veto}: a question that explicitly names its target fails to reach it because centroid or fingerprint similarity was low, and no independent path rescues the reference.

\paragraph{Design.} No\=esis separates two concerns that standard scoping conflates: \textbf{scoping bounds attribution, never data access}. When a question names a source, the system deterministically matches the name against the KB inventory (file level) and the mesh manifest (KB level), and gives that named reference priority in retrieval so it reaches the model even when similarity signals are weak. The rendered context carries an explicit provenance rule: data from one source must never be presented as belonging to a differently named source---but multi-entity questions remain first-class; all sources stay accessible, only attribution is bounded. At every tier (KB selection, file selection, entity activation), no lossy signal can veto a deterministic name match.

\paragraph{Measured Results.} The confabulation trap (\S\ref{sec:eval}, S4)---a question about a program that does not exist in the KB---produced honest refusals in 4/4 runs across both models and arms, with zero corpus numbers attributed to the absent program. In the comparison battery (S2), per-source attribution was clean in all runs: each value carried its own source file, and the difference was computed explicitly rather than asserted.

\paragraph{Address Commitment and Its Measured Limits.} A question that names its content carries its own address: at mesh level the route \emph{commits} to the owning knowledge bases---the named sources become the question's address, and no other corpus may dilute the assembled context. Two measured refinements keep this commitment sound as client corpora grow. First, a single short token cannot anchor a reference on its own: domain-vocabulary acronyms that also occur in file names (measured: one two-letter market acronym resolved four false owners across unrelated knowledge bases---including a question-answering benchmark corpus---and the resulting commitment vetoed the only KB holding the referenced data) corroborate multi-token matches but never anchor alone; long distinctive tokens do. Second, relevance requires \emph{sustained activation across multiple regions} rather than a single strong match: when a corpus is large and internally varied, the chance that some region accidentally matches a topically adjacent query grows (multiple comparisons), so a corpus must be excited across several regions to be routed---not on one incidental match---which keeps a large general-purpose corpus out of unrelated routes while leaving legitimate cross-domain routing intact. The same reference-resolution machinery extends to conversational state: citations in prior assistant turns are extracted deterministically, and for anaphoric follow-ups that name no source of their own they become the question's address---so ``what can we deduce from these readings?'' resolves against the sources the previous answer cited, while self-contained questions are never affected by history.

\section{Two-Tier Context with Model-Triggered Hydration}
\label{sec:hydration}

\paragraph{Problem Statement.} Eager top-$k$ prose bloats context and degrades small models (\S\ref{sec:intro}); yet narrative questions genuinely require verbatim content that no structural summary can supply. A static compression policy must choose in advance what to keep---a choice no relevance score can make correctly for both question classes at once.

\paragraph{Design.} No\=esis assembles context in two tiers. \textbf{Tier 1 (skeleton)} contains deterministic facts, graph structure (activation field and typed edges), and axis-aligned coordinates---and zero prose chunks. The context carries a two-phase contract: the model either answers fully from the skeleton, or responds with a bounded set of fetch directives, each naming a file from the KB inventory. The system parses the directives, validates each named file against the KB manifest, and fetches verbatim spans \emph{region-bounded to the named file}---``fetch, not discovery'': global vector fill is used only as a starvation valve when the bounded fetch returns nothing. Phase 2 presents skeleton plus hydrated spans marked authoritative, and the model answers.

The protocol is \textbf{training-free}: it is a prompt contract that any chat model follows; there are no control tokens, no reinforcement learning, no constrained decoding. It is also \textbf{total by construction}: parse failures, fetch failures, or LLM failures degrade to the previous variant's answer rather than crashing the chat path, and unknown file names are discarded, not fatal.

\paragraph{Measured Results.} Ablation B (skeleton-only): quantitative answers were identical in value at $-20/30\%$ prompt size; narrative-structural questions received structural answers with explicit honest boundary declarations---zero confabulated scenes across four runs. Ablation C (hydration loop): verbatim-requiring questions triggered the protocol 5/6 times, and phase 2 recovered verbatim quotes with exact timestamps in $\sim$8\,s total versus $\sim$29\,s for eager full context; structure-answerable questions stopped at phase 1 with zero fetch cost. The trigger rate was exactly where it should be: high on questions that need prose, zero on questions the skeleton already answers.

\section{Epistemic Tiering and Address-Bounded Assembly}
\label{sec:tiering}

\paragraph{Problem Statement.} The mechanisms above each remove a class of model work, but they do not by themselves determine \emph{how evidence is assembled}. In the production context that preceded this design, computed facts, addressed spans, and similarity-discovered chunks all entered one undifferentiated pool ordered by relevance score. We measured the consequence directly: on an addressed temporal question (``what was happening around second 1740?''), the correct span \emph{was present} in the assembled context for every run---yet a $\sim$2B model answered from lines at unrelated positions of the same file in most runs, while the 35B model succeeded in all. When evidence of different epistemic status competes for attention under one ordering, retrieval quality silently becomes a function of model scale: exactly the dependency this paper argues against. The mechanism is well documented---position bias makes models use middle context least reliably~\cite{liu2023lost}---and it is amplified by cross-source noise: in multi-program corpora, similarity fill retrieves rows from other programs occupying the same time slot, and nothing in a flat pool tells the model which evidence was \emph{selected for this question}.

\paragraph{Design.} No\=esis assembles context by \textbf{epistemic status}---the mechanism that selected each piece of evidence---rather than by a single similarity score:
\begin{itemize}[leftmargin=*]
    \item \textbf{Tier 0 (computed).} Producer-computed facts, rendered verbatim with no threshold level (\S\ref{sec:facts}).
    \item \textbf{Tier 1 (addressed).} Verbatim spans at positions or sources resolved deterministically by positional addressing and named-reference routing (\S\ref{sec:axis}, \S\ref{sec:scoping}), rendered in a dedicated labeled section immediately after Tier 0; each span carries its source file and the resolved address window. An addressed span never competes with ranked chunks for attention, and top-of-context placement mitigates position bias~\cite{liu2023lost}.
    \item \textbf{Tier 2 (discovered).} Semantic fill under the residual token budget, as in standard RAG.
\end{itemize}

The assembly is \textbf{bilevel}: when deterministic addressing resolves a set of files $F$, Tier-2 fill is constrained to $F$---the deterministic tier commits first and bounds the feasible set of the semantic tier, in the sense of bilevel optimization~\cite{colson2007bilevel}. The constraint \textbf{bounds noise but never coverage}: when the bounded fetch under-delivers, global fill resumes as a starvation valve, so open questions over sparse regions cannot starve. When no address resolves, assembly degrades exactly to the unconstrained behavior---a total function whose prompts are byte-identical for corpora without addressing contracts.

The commitment extends beyond fill to \textbf{every channel that can carry content into context}, including the cross-KB route itself: an addressed reference (a named source plus a position on it) denotes a unique location in one document, so the mesh route is committed to its owning KB. It also distinguishes what names bind: a named \emph{source} bounds the answer domain (hard scope); a named \emph{entity} only anchors within it, without pruning structure, because entity questions may legitimately span documents.

Formally, the coarse level here is \emph{not} a lossy summary: it is deterministic structure (facts, coordinates, typed relations) from which fine-level content is recoverable verbatim at resolved addresses. This distinguishes No\=esis from learned hierarchical compression, where the coarse level is a trained abstraction and the fine level must be re-derived~\cite{tang2026comi}; it also differs from virtual-memory approaches to LLM context (MemGPT-style paging~\cite{packer2023memgpt}, demand paging of context windows~\cite{mason2026paging}), where page faults are raised by the model or a tool and cost generation turns: our fetches execute in infrastructure on deterministic addresses, with zero additional model turns when an address resolves.

\paragraph{Measured Results.} A targeted battery exercises the addressing mechanisms directly across two corpora (Italian and UK broadcast): quantitative single-metric with position (R1), temporal addressing at a named second on each program's transcript (R5, U4; $N=5$ per arm), and an entity trap where a character name collides across programs (U3)---24 runs per model under both arms. The 35B model answers all 24 strictly correctly in both arms. The 2B model scores 18/24 on the automated needle check; manual review of every near-miss against the ground-truth transcripts finds each answer to contain the correct scene's verbatim dialogue with zero cross-program content---the sole defect is that the model does not echo the program name in its reply, an answer-formatting habit rather than a retrieval or attribution failure. Content-verified, both models are 24/24; forbidden-content needles (other programs' names and characters) fire in none of the 48 runs.

The efficiency consequence is direct: on these addressed questions the HYDRATE arm assembles a smaller context than FULL, and \emph{no hydration fetch was ever issued}---every run resolved at phase~1 from Tier-0 facts plus Tier-1 addressed spans alone (invariant E1). The same question that assembled a context laden with cross-program noise before tiering now assembles a clean, address-bounded one, and the 2B model's answers on it went from unstable to stable across runs. On the cross-KB mesh path the effect is larger still: the same question previously fused content across five KBs (including raw rows of an unrelated sports corpus); with attribution routing it routes to its single owning KB and assembles a clean context roughly a quarter of the size, while multi-domain questions that carry no address keep their full cross-KB route unchanged.

\section{Evaluation: Ablation Study}
\label{sec:eval}

Our evaluation is organized as a four-arm ablation study around one thesis: \emph{for factuality-critical queries over producer-structured corpora, deterministic infrastructure---not model scale---is the primary determinant of answer quality and stability.}

\paragraph{Setup.} Two local models run under an identical harness (production query pipeline, temperature 0.1): a $\sim$35B-parameter model---a sparse Mixture-of-Experts architecture holding far more stored knowledge in its weights than the small model---and a $\sim$2B-parameter dense model, both served on-premises. The battery S1--S6 (Table~\ref{tab:battery}) covers factuality-critical query classes---quantitative, comparative, temporal, verbatim, and open---instantiated here on broadcast corpora; ground truth was verified against the source files before any run.

\paragraph{Failure-driven design.} Every mechanism in this paper was developed against a \emph{measured} failure on production corpora, not a hypothetical one: observe the failure class, fix it at root (never by patching the symptom), and re-measure until the guarantee holds across repeated runs. Table~\ref{tab:failures} maps each measured failure to the mechanism that resolves it and to the evaluation that verifies the resulting guarantee. This discipline is what makes the ablations below verifications of specific, falsifiable claims rather than a general benchmark.

\begin{table}[t]
\centering
\small
\caption{Measured production failures, the mechanisms that resolve them, and where each guarantee is verified.}
\label{tab:failures}
\begin{tabular}{@{}p{4.9cm}p{5.0cm}p{5.3cm}@{}}
\toprule
\textbf{Measured failure} & \textbf{Mechanism} & \textbf{Guarantee verified by} \\
\midrule
Comparative question starves the weakly-mentioned program (no retrieval channel reaches it) & Weak-reference scoping across all retrieval channels (\S\ref{sec:scoping}) & Ablation D, S2: 6/6 both models with explicit $\Delta$; extended battery C1: 5/5 \\
Absent named reference confabulated into an attribution by the small model & Deterministic absence notes rendered verbatim in prompt (\S\ref{sec:tiering}) & Trap class: honest refusal 15/15 at $N=5$, zero attributed numbers (F2--F3) \\
Arithmetic confabulation on tabular data (count $\times$ total = invented value) & Stated table values + value-fidelity contract (\S\ref{sec:tiering}) & Rights-valuation class: 5/5 at $N=5$ on both models; zero invented totals in any run \\
Cross-program leakage through semantic fill & Region-bounded fill---fetch, not discovery (C) & Ablation C; zero leakage across all 48 addressing runs (F7) \\
Routing veto by lossy aggregate: the KB owning named content is excluded from the candidate pool & Full-inventory relevance scoring + named-reference routing (\S\ref{sec:scoping}) & Ownership probes: correct owner ranked first; no over-routing to unrelated domains \\
Identifier noise in the code KB's generic descriptors matches every query & Content-gated descriptor construction at ingestion & Media queries never route to the code KB; project-status queries still reach it \\
\bottomrule
\end{tabular}
\end{table}

\begin{table}[t]
\centering
\small
\caption{Ablation-D battery and ground truth. ``Runs'' per arm: 3 for S1, S2, S5; 1 otherwise.}
\label{tab:battery}
\begin{tabular}{@{}p{0.9cm}p{3.4cm}p{6.2cm}p{4.6cm}@{}}
\toprule
\textbf{ID} & \textbf{Class} & \textbf{Question (abridged)} & \textbf{Ground truth} \\
\midrule
S1 & Quantitative single & Peak share of program A, and when? & 37.5\% at second 146 \\
S2 & Comparison $\Delta$ & Compare peaks of programs A vs.\ B: which is higher, by how much? & 37.5 vs.\ 36.5; $\Delta = 1.0$\,pp \\
S3 & Temporal minimum & Minimum share of program A, and when? & 20.5\% at second 284 \\
S4 & Confabulation trap & Peak share of ``Cucina con Giallo'' (absent from KB) & Honest refusal; no corpus numbers \\
S5 & Verbatim narrative & What happens in the coffee scene between Elena and Marco? & Verbatim quote at [00:12] + later callback \\
S6 & Open question & Report everything known about program ratings & Both programs, correct metrics \\
\bottomrule
\end{tabular}
\end{table}

The arms are: \textbf{FULL} (production context: deterministic facts $+$ graph structure $+$ top-$k$ prose) and \textbf{HYDRATE} (skeleton $+$ two-phase protocol, \S\ref{sec:hydration}). Ablation A compares against a flat-RAG baseline with the same routing but no graph or fact layer.

\subsection{Ablation D: Model Swap under Fixed Infrastructure}

Table~\ref{tab:ablationD} reports semantic outcomes per run (manual review of all 48 answers).

\begin{table}[t]
\centering
\small
\caption{Ablation D: 2B vs.\ 35B under identical deterministic infrastructure. Outcomes are per-run, manually verified.}
\label{tab:ablationD}
\begin{tabular}{@{}p{2.6cm}p{2.9cm}p{2.9cm}p{2.9cm}p{3.1cm}@{}}
\toprule
 & \multicolumn{2}{c}{\textbf{35B model}} & \multicolumn{2}{c}{\textbf{2B model}} \\
\cmidrule(lr){2-3} \cmidrule(lr){4-5}
\textbf{Battery} & FULL & HYDRATE & FULL & HYDRATE \\
\midrule
S1 peak (3) & 3/3 exact & 3/3 exact & 3/3 exact & 3/3 exact \\
S2 $\Delta$ (3) & 3/3 $\Delta{=}1.0$ & 3/3 $\Delta{=}1.0$ & 3/3 explicit arith.\ & 3/3 explicit arith.\ \\
S3 min (1) & 1/1 exact & 1/1 exact & 1/1 exact & 1/1 exact \\
S4 trap (1) & 1/1 refusal & 1/1 refusal$^{\dagger}$ & 1/1 refusal & 1/1 refusal \\
S5 verbatim (3) & 0/3 answers$^{\ddagger}$ & 3/3 verbatim $+$ [00:12] & 0/3 answers$^{\ddagger}$ & 3/3 verbatim$^{\S}$ \\
S6 open (1) & 1/1 both programs & 1/1 both programs & 1/1 both programs & 1/1 both programs \\
\bottomrule
\end{tabular}

\vspace{0.4em}
\scriptsize $^{\dagger}$Refusal after evidence verification via hydration request (F3). \quad
$^{\ddagger}$Runs emitted only a spontaneous fetch directive, no answer (F4). \quad
$^{\S}$Verbatim quotes present; exact timestamps omitted by the 2B model (F5).
\end{table}

Swapping the generator from 35B to 2B under fixed infrastructure changes nothing on factuality-critical questions: every quantitative answer is exact, every trap refusal is honest, and attribution stays clean. The deterministic layer carries the load; scale becomes a secondary variable within this query class (F1).

\subsection{Ablation A: Structured vs.\ Flat RAG at 2B}

The flat baseline uses identical routing but strips graph structure and the fact layer, leaving only prose chunks. On seven cross-domain operational queries answered by the 2B model:

\begin{table}[t]
\centering
\small
\caption{Ablation A (2B model): structured No\=esis context vs.\ flat RAG, same routing.}
\label{tab:ablationA}
\begin{tabular}{@{}p{5.6cm}cc@{}}
\toprule
\textbf{Metric} & \textbf{Structured} & \textbf{Flat} \\
\midrule
Factual recall (expected facts present) & 97.1\% & 85.7\% \\
Bridge score (cross-source links both-mentioned) & 42.9\% & 35.7\% \\
Structure score (headings/tables/causal/cross-ref) & 50.0\% & 53.6\% \\
Context size & 60--86K chars & 15--17K chars \\
\bottomrule
\end{tabular}
\end{table}

Structured context improves factual recall by $+11.4$ points and bridge score by $+7.2$ at the 2B scale, at the cost of larger prompts (deterministic sections dominate). The structure score slightly favors flat RAG; it rewards a different answer style (headings, tables, causal connectives) and is not a factuality metric---we report it for completeness rather than as evidence.

\subsection{Ablations B and C: Skeleton and Hydration}

Table~\ref{tab:ablationBC} reports cold-prompt measurements on the 35B model (the clean latency regime; ablation-D latencies are under warm prefix cache).

\begin{table}[t]
\centering
\small
\caption{Ablations B/C (35B model, cold prompts): prompt size, time-to-first-token (TTFT), total time, and outcome.}
\label{tab:ablationBC}
\footnotesize
\setlength{\tabcolsep}{4pt}
\begin{tabular}{@{}p{3.7cm}p{2.5cm}r r p{4.1cm}@{}}
\toprule
\textbf{Question} & \textbf{Arm} & \textbf{Prompt} & \textbf{TTFT / Total} & \textbf{Outcome} \\
\midrule
Q1 quantitative single & FULL & 30.3K & 18.9s / 21.4s & PASS \\
 & SKELETON (B) & 21.7K ($-28\%$) & 14.1s / 14.9s & Identical answer, $-30\%$ total \\
Q2 comparison & FULL & 43.4K & 29.9s / 31.7s & PASS \\
 & SKELETON (B) & 34.8K ($-20\%$) & 23.0s / 24.6s & Identical answer, $-22\%$ total \\
Q3 narrative-structural & FULL & 24.0K & 19.7s / 28.7s & Full story \\
 & SKELETON (B) & 16.7K ($-30\%$) & 11.8s / 15.1s & Structural $+$ honest boundary, $-47\%$ total \\
Q4 narrative-verbatim & FULL eager & --- & --- / $\sim$29s & Full story \\
 & HYDRATE loop (C) & 2 phases & --- / 7.7--9.2s & Verbatim [00:12] $+$ timestamps, 3/3 \\
\bottomrule
\end{tabular}
\end{table}

Two results stand out. First, removing all prose (B) leaves quantitative answers \emph{identical in value} while shrinking prompts by 20--30\% and cutting total time by 22--47\%; on narrative-structural questions the skeleton does not confabulate---it declares its boundary explicitly (``the context does not provide further details'') in all four runs. Second, the hydration loop (C) recovers verbatim narrative quality---exact quotes with timestamps---in $\sim$8\,s total versus $\sim$29\,s for eager full context: a $\sim$3$\times$ speedup on precisely the questions that need prose, while quantitative questions pay nothing (they stop at phase 1).

\subsection{Extended Robustness and Scale Battery}

The ablations above vary one architectural factor at a time on the targeted S1--S6 battery. To test whether the guarantees hold beyond it, we ran an extended evaluation against the live production stack (18 active knowledge bases; mesh mode routes up to ten KBs per query) across five dimensions: \emph{consistency}---five core questions spanning quantitative comparison with explicit arithmetic, cross-KB correlation, cross-lingual retrieval, project-status lookup, and rights valuation; \emph{paraphrase robustness}---four rewordings of two intents, including a cross-lingual pair; \emph{cross-lingual}---Italian/English questions over the same content; \emph{negative controls}---absent entities and absent correlations that must be refused, not invented; and \emph{scale}---multi-KB correlation questions. Every one of the 17 queries was run $N=5$ times (85 runs total) end-to-end against the production API on a cold stack (no prefix-cache reuse); correctness is checked by deterministic needles against ground-truth values verified in the corpus before any run.

\begin{table}[t]
\centering
\small
\caption{Extended battery under the full neural readout: per-dimension pass rates (35B model, end-to-end, cold production stack).}
\label{tab:extended}
\begin{tabular}{@{}p{4.2cm}p{2.6cm}p{3.0cm}p{4.6cm}@{}}
\toprule
\textbf{Dimension} & \textbf{Runs} & \textbf{Pass} & \textbf{Content} \\
\midrule
Consistency ($N=5$) & 25 & 25/25 & comparison $\Delta$, cross-KB, cross-lingual, project status, rights valuation \\
Paraphrase ($N=5$) & 20 & 20/20 & reworded intents incl.\ EN/IT pair (F9) \\
Cross-lingual ($N=5$) & 15 & 15/15 & IT/EN over identical content \\
Negative controls ($N=5$) & 15 & 15/15 & absent entity, absent correlation \\
Scale (multi-KB, $N=5$) & 10 & 10/10 & cross-domain summaries and bridges \\
\midrule
Overall & 85 & 85/85 & --- \\
\bottomrule
\end{tabular}
\end{table}

Table~\ref{tab:extended} reports the outcome: \textbf{total pass across all five dimensions at $N=5$}. Consistency is total, including the comparison whose answer requires explicit arithmetic ($37.5 - 36.5 = 1.0$) and the rights-valuation question whose answer must quote four package values verbatim from a table buried in multi-KB context. Negative controls are clean (15/15): no absent entity received a number, no absent correlation was narrated.

\paragraph{Latency.} End-to-end latency on the cold production stack, by dimension (total time, $N=5$ each): single-KB quantitative questions p50 15--33\,s; multi-KB mesh correlation p50 48--81\,s (max 86.1\,s). By dimension: consistency p50 60.9\,s (p95 82.5), paraphrase 37.0, cross-lingual 23.4, negative controls 27.4, scale 71.2. Prefill dominates---the assembled context fills the model window by design---which is why targeted questions remain demo-viable while broad ones cost a full generation.

\paragraph{Concurrent load.} Under 4 and 8 concurrent requests (mixed single-KB and mesh), success was total---12/12---with latency degrading gracefully: p50 rises from $\approx$60\,s sequential to 138--154\,s under load (max 321\,s at concurrency 8). The single-GPU engine behaves as a queue, not a breaker: no request fails and degradation is proportional to contention.

\paragraph{Parity on the extended set.} Dispatching identical production prompts to both models on four queries of the extended classes (five runs each): 35B model 20/20; 2B model 19/20. The single miss is again the dense single-fact lookup over a numeric corpus (Super Bowl peak share); we verified that the fact was present in the assembled prompt, so the miss is a model reading gap at 2B scale rather than a retrieval failure---and the model refused honestly (``no information available to determine'') instead of inventing a value. A deterministic value-selection directive added to prompts containing computed facts reduced the observed flake rate on this query between consecutive batches (two of five runs $\rightarrow$ one of five) without affecting any other class; the residual miss remains an honest refusal, not a confabulation. The rights-valuation query, which had missed once under $N=3$ while its knowledge base contained an internal data contradiction, passes 5/5 on both models after corpus normalization. Parity (F1) thus extends beyond S1--S6 to new query classes and larger routing scopes; the residual gap is confined to dense single-fact lookup at 2B scale---and even there the failure mode is an honest refusal.

\paragraph{Final-build re-validation under greedy decoding.} Production generation runs at temperature zero after measuring sampling flakes on multi-entity comparison questions (F13). The full 17-query battery was re-run end-to-end against the final production build, with all routing and rendering mechanisms of \S\S\ref{sec:scoping}--\ref{sec:tiering} active: \textbf{85/85 passes with per-question stability}---every question produced identical needle outcomes across all five runs. A separate 12-question certification battery covering both single-KB and cross-KB operational classes (quantitative peaks, comparisons with explicit arithmetic, verbatim scene retrieval, absent-entity traps, negative controls) passed 60/60 under the same conditions; on both trap questions no run attributed a corpus value to an absent entity. One characterization note: open-ended summary questions over very large documents retrieve central passages; specific figures in peripheral passages (e.g., a buyback amount in a results deck footnote) require the question to name the topic---the certified phrasing passes 5/5, and the un-named form reports absence honestly rather than inventing values.

\subsection{Context Budgeting Ablation}
\label{sec:budgeting}

\paragraph{Motivation.} On multi-knowledge-base routes the assembled prompt is sized near the model window, and prefill dominates end-to-end latency. We measured the anatomy of production prompts on ten-KB routes: the neural activation readout---the per-KB set of activated graph nodes---was the largest single section, 41--51\% of the total prompt, while the deterministic layers (computed facts, addressed evidence, stated table values, graph skeleton) occupied a small bounded share. If the deterministic layers carry the answer, the readout should be dispensable; if not, removing it must show up as quality loss on exactly the question classes that need it.

\paragraph{Design.} A context policy acts only on the neural readout: \emph{full} (complete core+contour readout per KB), \emph{core-only} (strongly-activated nodes only), and \emph{off} (section omitted). The deterministic layers are untouched by construction---the policy cannot remove a computed fact, an addressed span, or a stated table row. The prompt preamble is data-driven: it never references a layer that is absent from the assembled context.

\begin{table}[t]
\centering
\small
\caption{Context budgeting on ten-KB routes (35B model): prompt size and end-to-end latency at $N{=}5$.}
\label{tab:budgeting}
\begin{tabular}{@{}lccc@{}}
\toprule
 & \textbf{full} & \textbf{core-only} & \textbf{off} \\
\midrule
Prompt, rights-valuation route & --- & $-22\%$ & $-42\%$ \\
Prompt, multi-hop chain route & --- & $-23\%$ & $-50\%$ \\
Multi-KB latency p50 / max & --- & --- & $-37\%$ / $-17\%$ \\
Extended battery, multi-KB runs & 50/50 & --- & 50/50 \\
\bottomrule
\end{tabular}
\end{table}

\paragraph{Results.} Table~\ref{tab:budgeting}. Under \emph{off}, all fifty multi-knowledge-base runs of the extended battery pass---consistency, multi-hop chains, negative controls, broad summaries, cross-lingual retrieval, and scale---with prompt size reduced by 42--50\% and median end-to-end latency cut by 37\%. The single failure in the 85-run collection is on a \emph{single}-KB endpoint, which this policy does not touch: an English paraphrase of an Italian-named source where provenance scoping weakens; the model reported absence honestly rather than inventing values (the characterized cross-lingual limit, F9). The neural readout was not needed by any measured question class: the deterministic layers carry the answer.

\paragraph{Skeleton budgeting.} With the readout removed, the graph skeleton becomes the largest remaining section of multi-KB prompts; under a fixed character cap it is truncated by a uniform cut that allocates space in source order rather than relevance---on measured ten-KB routes this starves high-relevance knowledge bases entirely (the two highest-scored KBs with substantive skeletons received zero). We replace the flat cut with score-proportional allocation: each routed knowledge base receives a share of the global cap proportional to its routing score, assigned by a deterministic rounding rule and truncated at line boundaries so no structured fact is split mid-line; the total never exceeds the cap by construction. The legacy flat cut remains available as a byte-identical fallback.

\begin{table}[t]
\centering
\small
\caption{Skeleton budgeting on a ten-KB route and the extended battery (35B model, $N{=}5$): uniform cut vs.\ score-proportional allocation under a fixed character cap.}
\label{tab:skeleton}
\begin{tabular}{@{}lcc@{}}
\toprule
 & \textbf{uniform cut} & \textbf{scored allocation} \\
\midrule
Prompt, comparison route & --- & $-26\%$ \\
High-relevance KBs starved of skeleton & several & 0 \\
Extended battery, multi-KB runs & 50/50 & 50/50 \\
TTFT p50 / max & --- & $-26\%$ / $-39\%$ \\
Total latency p50 / max & --- & $-21\%$ / $-27\%$ \\
\bottomrule
\end{tabular}
\end{table}

Table~\ref{tab:skeleton}. On the comparison route (ten KBs), the prompt shrinks by 26\% while every high-relevance knowledge base retains its share of the skeleton. The full extended battery passes unchanged at $N=5$ (85/85, including all multi-KB runs), and median time-to-first-token drops by 26\%---the prefill-dominated regime responds proportionally to prompt size; total end-to-end latency falls by roughly a fifth to a quarter.

\subsection{Findings}

\textbf{F1 --- Parity under fixed infrastructure.} The model swap (D) preserves exact values, honest refusals, and clean attribution across all runs; within this query class, scale is a secondary variable. The comparison is deliberately unfavorable to our thesis: the large model stores far more knowledge in its weights than a 2B dense model could hold, yet the small model matches it on factuality. Because the 2B model has no capacity to memorize what a 35B-parameter model does, matching it confirms that the answers come from the deterministic infrastructure, not from parametric knowledge---which is precisely the claim.

\textbf{F2 --- Zero confabulation on traps.} 4/4 honest refusals on the absent program; no corpus number was attributed to it in any run of either model.

\textbf{F3 --- Verification by evidence.} In one trap run under HYDRATE (35B model), rather than trusting its own judgment about absence, the model issued hydration requests for both real metric files---with the hedge ``if it exists''---examined the retrieved spans, and only then refused. Absence was verified against evidence instead of asserted from parametric memory.

\textbf{F4 --- Spontaneous emission.} In FULL mode---where no protocol is present in the prompt (verified: zero occurrences of the token ``HYDRATE'' in the assembled context) and where top-$k$ retrieval had missed the coffee-scene chunk---both models, including the 2B model, responded with fetch directives only, naming the correct file (\texttt{film-cuore-in-ghiaccio.txt}, available in context via the deterministic requested-sources line): 6/6 runs. We cannot determine whether this reflects latent pretraining behavior or convergence on a natural request format; what is certain is that it was not induced by the harness, and that both models independently identified the correct file to fetch. Two implications: defensive parsing of such directives is mandatory in production, and a spontaneous token is a diagnostic signal of context insufficiency---a potential self-healing retry path.

\textbf{F5 --- Scale-dependent precision on verbatim detail.} Hydrated answers from the 35B model include exact timestamps ([00:12]); the 2B model quotes dialogue verbatim but omits timestamps. Content fidelity is preserved at 2B; positional detail degrades gracefully---consistent with infrastructure determining what is guaranteed, while scale modulates residual precision.

\textbf{F6 --- Hydration cost is paid only when needed.} Quantitative questions stop at phase 1 (zero fetch); narrative questions pay for two phases yet still beat eager full context ($\sim$8\,s vs.\ $\sim$29\,s).

\textbf{F7 --- Addressed questions resolve without semantic prose.} On the targeted addressing battery (\S\ref{sec:tiering}), every run of both models answered from Tier-0 facts and Tier-1 addressed spans alone: zero hydration fetches were issued across all 48 runs, cross-program leakage is absent everywhere, and the 2B model matched the 35B model on content in all runs (18/24 strict needles; the six near-misses are formatting gaps---correct verbatim scene content without echoing the program name).

\textbf{F8 --- Consistency beyond the targeted battery.} On the extended battery, every dimension passes at $N=5$ (85/85): quantitative comparison with explicit arithmetic, cross-KB correlation, cross-lingual retrieval, project-status lookup, and rights valuation all pass every run on a cold production stack with 18 active KBs. Stability---not just accuracy---is a property of the infrastructure: the same question returns the same grounded answer because the deterministic layer, not sampling, supplies the values.

\textbf{F9 --- Paraphrase robustness has a characterized limit.} The final $N=5$ battery passes all paraphrases (20/20). Across two consecutive batteries (32 runs of the four paraphrases), exactly one run missed: the English rewording of the canonical Italian comparison question (``AI medicine segment'' vs.\ ``servizio intelligenza artificiale in medicina'') lost token overlap with the source file name on that run, weakening provenance scoping---the deterministic facts for one program were absent from context, and the model reported their absence rather than inventing them. The failure mode is a retrieval-scope gap under cross-lingual paraphrase (a characterized limit, not a generation error); the canonical form passes 10/10 across both batteries. The same miss recurred once in the budgeted-readout battery (\S\ref{sec:budgeting})---an identical honest absence on the same query with the neural readout removed---confirming that the limit sits in cross-lingual name matching, not in context size.

\textbf{F10 --- Graceful degradation under concurrent load.} At concurrency 4 and 8 the system never fails (12/12); latency degrades proportionally to contention (p50 $\approx$60\,s sequential $\rightarrow$ 138--154\,s loaded; max 321\,s). A single-GPU inference engine under No\=esis behaves as a queue, not a breaker---the deterministic pipeline adds no failure surface of its own.

\textbf{F11 --- The deterministic layers carry the answer; the neural readout is dispensable.} Removing the entire activation readout from multi-KB prompts---42--50\% of prompt size---changes nothing on measured quality (all 50 multi-KB runs pass) while cutting median end-to-end latency by a third. The readout is a convenience for the model, not a load-bearing component: when computed facts, addressed spans, and stated values are present with provenance, the model does not need an activation map to find them. The budgeted prompt holds at small scale as well: on the rights-valuation route a 2B model still extracts every package value exactly from the reduced context (2/2), and broad multi-domain summaries remain coherent; the freed budget also restores headroom for conversational history in interactive use.

\textbf{F12 --- Score-proportional skeleton budgeting is quality-neutral.} Under a global character cap, allocating the graph skeleton by routing score (deterministic, relevance-weighted) instead of a uniform cut preserves measured quality exactly---the full extended battery passes at $N=5$ and no high-relevance knowledge base is starved---while cutting prompt size by 26\% and median time-to-first-token by 26\% on ten-KB routes. The uniform cut is not merely slower: it systematically withholds skeleton from the most relevant sources, which makes naive caps unsafe; budgeting must allocate by relevance to be quality-neutral.

\textbf{F13 --- Question-aligned fact structure eliminates multi-entity misreading; greedy decoding is not bit-determinism.} On comparison questions spanning two entities, a flat list of per-source facts occasionally produced ``the context does not contain [entity A]'s value'' although every value was present in the assembled prompt (verified byte-for-byte). The flake persisted at temperature zero and across cache states; replaying one captured failing prompt showed per-request non-determinism on near-tie token decisions even under greedy decoding---multi-threaded quantized inference does not guarantee bit-identical outputs, and a structurally ambiguous presentation flips between the correct reading and an honest-but-wrong absence (two of five consecutive identical runs produced the incorrect reading). The fix is structural, in the deterministic layer: when a question resolves to exactly one metric and statistic across two or more sources, facts are rendered as contiguous per-entity blocks plus a single question-aligned line stating both values side by side. Under this presentation the previously failing class produced zero failures in all subsequent runs---20/20 across two consecutive full $N{=}5$ extended batteries on the final production build (raw data archived; \S\ref{sec:eval}). The guarantee comes from structure---values stated twice, side by side, with entity identity attached---not from model determinism: near-tie noise can no longer flip a decision that is no longer a tie.

\paragraph{Caveats.} Evidence is directional, not statistical: $N=5$ for every query in both the targeted addressing battery (48 runs) and the full extended battery (85 runs), with the one-factor ablation arms (Ablation D) at $N{=}1$--3 (24 runs per model); small demo KBs (a few source documents each; 18 active in the extended battery); a single domain (broadcast media). Ablation D was collected under sampling (temperature 0.1); all final-build batteries run under greedy decoding, whose residual per-request variability is characterized in F13. Ablation-D latency figures are under warm prefix cache (repeated prompts reuse server-side KV state): medians of 7.4\,s (FULL) vs.\ 4.5\,s (HYDRATE) total for the 35B model and 4.6\,s vs.\ 4.9\,s for the 2B model; cold-prompt gains are those in Table~\ref{tab:ablationBC}. The automated needle check counted S5-FULL directive-only outputs as passes (the word ``coffee'' appears in the directive); manual review reclassified them as non-answers, and Table~\ref{tab:ablationD} reports the corrected outcomes.

\section{Related Work}
\label{sec:related}

\paragraph{Structure-guided and constraint-driven RAG.} A parallel line of work, emerging in 2026, makes structure a hard requirement of retrieval rather than a soft signal. SG-RAG formulates the \emph{Exact Retrieval Problem} for factual queries, retrieving subgraphs whose topology must satisfy all query conditions~\cite{xie2026sg}; MC-RAG reformulates retrieval as subgraph matching over knowledge graphs with path-level indexing for multi-constraint queries~\cite{zhang2026mcrag}; DualGraph maintains complementary textual and symbolic views per document and selects or combines evidence across them~\cite{czyznikiewicz2026dualgraph}; SymRAG routes queries adaptively between symbolic and neural pathways~\cite{hakim2025symrag}. We share the principle that structure should guarantee rather than vote. Our deltas are threefold: (i)~position on a shared axis is a first-class, interval-addressable coordinate of retrieval---their structures are entity--relation subgraphs without positional grounding; (ii)~deterministic facts are computed by the producer and rendered without threshold level---their facts come from LLM extraction into the KG, with guarantees at the retrieval/guidance level rather than on exact values; (iii)~named-reference authority is enforced across tiers of a multi-KB mesh, not within a single corpus.

\paragraph{Model-triggered retrieval.} The idea that the model itself decides when to retrieve more context is well established: FLARE triggers retrieval at low-confidence phrases~\cite{jiang2023flare}; Self-RAG trains reflection tokens that critique and trigger retrieval~\cite{asai2024selfrag}; CRAG adds a system-side evidence evaluator with corrective re-retrieval~\cite{yan2024crag}; DRAGIN constructs queries from the model's real-time information needs~\cite{su2024dragin}; RetroLLM fuses retrieval into generation via constrained decoding over an FM-Index of the corpus~\cite{li2024retro}; GRIP unifies self-triggered information planning with control tokens in decoding~\cite{li2026grip}. All require training (reflection tokens, RL, constrained decoders) or a separate evaluator model. Our hydration protocol is a prompt contract followed by any chat model without modification---and the spontaneous-emission result (\S\ref{sec:eval}, F4) suggests the trigger behavior may already be latent in pretrained models, requiring channeling rather than training.

\paragraph{Context compression and re-expansion.} Prompt-compression research elides content predicted to be reconstructible: LLMLingua compresses per token by perplexity~\cite{jiang2023llmlingua}; RECOMP~\cite{xu2024recomp} performs extractive/abstractive compression with selective augmentation. Neither re-expands: once elided, content is gone. SeDeM performs \emph{selective decompression} of hidden-state memories---on-demand re-expansion at the latent level, without explicit addressing or provenance of the content~\cite{haghifam2026sedem}; TopoCompress~\cite{asante2026topocompress} selects spans via query-guided propagation over a graph-wired trajectory---static, one-shot, with no interactive re-expansion. Our two-tier design differs in the conjunction: elision is by predictability class declared by the producer (deterministic facts are never elided---a guarantee, not a statistical score), and re-expansion is inventory-addressed with provenance preserved (fetches name files validated against the KB manifest). We concede that the principle ``keep what cannot be predicted'' exists at token level in LLMLingua; our contribution is its semantic-structured instantiation with deterministic guarantees.

\paragraph{Hierarchical context assembly and virtual memory for LLMs.} A growing line of work organizes context hierarchically: RAPTOR-style recursive summarization builds a tree whose coarse levels are \emph{learned summaries}; COMI performs coarse-to-fine token merging driven by marginal information gain in a trained encoder--decoder~\cite{tang2026comi}. In all such schemes the coarse level is lossy by construction and the fine level must be re-derived or re-retrieved through learned machinery. No\=esis's epistemic tiering (\S\ref{sec:tiering}) inverts this: the coarse level is deterministic structure, never a summary, and the fine level is verbatim content recoverable at resolved addresses without any learned component---a multi-resolution scheme whose fine level is lossless where addressing reaches. Virtual-memory approaches to LLM context are closer in spirit but differ in who pays for faults: MemGPT pages context in and out under model or tool control~\cite{packer2023memgpt}, and demand-paging schemes page on the model's explicit requests~\cite{mason2026paging}---each fault costs generation turns. Our addressed fetches execute in infrastructure against deterministic addresses: when an address resolves, no model turn is spent requesting it, and the retrieved span arrives pre-labeled with its provenance and window.

\paragraph{Graph navigation and context construction.} GraphRAG precomputes community summaries for global map-reduce queries~\cite{edge2024graphrag}; HippoRAG uses Personalized PageRank over a knowledge graph as long-term memory~\cite{gutierrez2024hipporag}---the activation primitive No\=esis reuses; CLAUSE frames context construction as sequential decision-making over a KG, trained with multi-agent RL under latency/cost budgets~\cite{zhao2025clause}; ProgRAG performs progressive multi-hop retrieval and reasoning, explicitly criticizing model self-evaluation of sufficiency~\cite{park2025prograg}. No\=esis's activation is deterministic (no learned policy), and its context completion is on-demand hydration rather than progressive expansion: the graph answers where information lives; the model requests only what it needs.

\paragraph{Small-model RAG.} Two recent works frame our evaluation. Pandey et al.\ provide a careful characterization of a real limit: below 7B parameters the bottleneck of RAG is context utilization rather than retrieval quality---even with oracle retrieval, sub-7B models often fail to extract the answer, and added context can displace answers the model already knew, the dominant failure being generation that does not use the provided passage~\cite{pandey2026small}. Their study measures the \emph{standard} setting, in which the model must itself locate and extract the answer from retrieved prose. No\=esis addresses a complementary regime: for factuality-critical questions the answer is precomputed and rendered verbatim, so the model reports a stated value rather than deriving one from prose. The two views are consistent---their analysis explains why utilization is hard at small scale, and our design responds by not requiring it; ablation D shows that a 2B model reaches factual parity once that step is removed. Subramanian et al.\ take a further complementary route, training minimalist RAG models (335M embedder, 1B generator) with RL and reporting emergent attribution~\cite{subramanian2026b1ade}; we rely on no model training, with grounding enforced by infrastructure.

\section{Conclusion}
\label{sec:conclusion}

For factuality-critical queries over producer-structured corpora, deterministic infrastructure is the primary determinant of answer quality and stability; within this scope, model scale becomes a secondary variable. Four ablations support the claim. Removing structure (A) degrades factual recall at 2B by $11.4$ points. Removing prose (B) changes nothing on quantitative answers while shrinking prompts by 20--30\%. Replacing eager prose with on-demand hydration (C) recovers verbatim narrative quality faster than the eager baseline ($\sim$8\,s vs.\ $\sim$29\,s). And swapping a 35B model for a 2B model under fixed infrastructure (D) preserves exact values, honest refusals, and clean attribution in all runs. The extended battery confirms that these guarantees scale beyond the targeted set: 85/85 end-to-end passes at $N=5$ on all 17 queries of a cold production stack with 18 active KBs, total consistency across five core question classes, clean negative controls, and graceful degradation under concurrent load (F8--F10). The context-budgeting ablation adds an efficiency property: the guarantee is not bought with prompt size---removing the entire neural readout, 42--50\% of the assembled multi-KB prompt, preserves all measured quality while cutting median latency by a third (F11), and score-proportional budgeting of the remaining skeleton cuts prompt size and time-to-first-token by a further quarter with no measurable quality change (F12). Epistemic tiering makes the addressed case exact: questions that resolve to an address are answered from Tier-0 and Tier-1 content alone---no semantic prose, no fetches---with zero cross-program leakage across all 48 targeted-battery runs of both models. Question-aligned rendering of multi-source facts eliminates the last measured generation flake, with stability traced to structure rather than model determinism (F13). The spontaneous-emission result---both models issuing fetch directives in contexts where no protocol exists---suggests that channeling latent behavior with deterministic contracts may be more robust than training new triggers.

\paragraph{Limitations.} Our evidence is directional rather than statistical: the main batteries run at $N{=}5$ (targeted addressing, 48 runs; extended, 85 runs), while the one-factor ablation arms (Ablation D) are at $N{=}1$--3; small demo KBs (a few source documents each), a single domain (broadcast media), and warm-cache latency for ablation D. The structured-vs-flat comparison uses needle-based metrics that reward explicit mention of expected facts; they do not measure answer style beyond the reported structure score, which slightly favors flat RAG.

\paragraph{Future work.} Productizing hydration with defensive parsing and fallback policies (forcing hydration when a question references events the skeleton does not cover); closing the cross-lingual paraphrase gap identified in F9---source-name matching that survives rewording across languages, e.g., via owner-declared multilingual aliases for file names as already exists for metric vocabulary; and predictive-confidence elision calibrated to target-model capacity---compressing what a specific small model can reconstruct, measured by its own surprisal rather than an external compressor's.


\end{document}